\pdfoutput=1
\documentclass{nature}
\usepackage{graphicx} 
\usepackage{cite}
\usepackage{epsfig} 
\usepackage{amsmath}
\usepackage{xcolor}
\usepackage{soul}
\usepackage{ulem}
\usepackage{caption}
\usepackage{xr}
\usepackage{lineno}

\title{Slow light silicon modulator beyond 110 GHz bandwidth}

\author{Changhao Han$^{1,7}$, Zhao Zheng$^{1,7}$, Haowen Shu$^{1,2{*}}$, Ming Jin$^{1}$, Jun Qin$^{3}$, Ruixuan Chen$^{1}$, Yuansheng Tao$^{1}$, Bitao Shen$^{1}$, Bowen Bai$^{1}$, Fenghe Yang$^{4}$, Yimeng Wang$^{1}$, Haoyu Wang$^{1}$, Feifan Wang$^{1}$, Zixuan Zhang$^{1}$, Shaohua Yu$^{1,2}$, Chao Peng$^{1,2,5\ddagger}$ and Xingjun Wang$^{1,2,5,6\dagger}$}
\begin{document}
\maketitle

\begin{affiliations}
\item State Key Laboratory of Advanced Optical Communication Systems and Networks, School of Electronics, Peking University, Beijing 100871, China
\item Peng Cheng Laboratory, Shenzhen 518055, China
\item Key Laboratory of Information and Communication Systems, Ministry of Information Industry, Beijing Information Science and Technology University, Beijing 100192, China
\item Zhang Jiang Laboratory, Shanghai 201210, China
\item Frontiers Science Center for Nano-optoelectronics, Peking University, Beijing 100871, China
\item Peking University Yangtze Delta Institute of Optoelectronics, Nantong 226010, China
\item These authors contributed equally to this work
\end{affiliations}

\noindent Corresponding Author:\\ $^{*}$haowenshu@pku.edu.cn, $^\ddagger$pengchao@pku.edu.cn, $^\dagger$xjwang@pku.edu.cn

\vspace{10pt}

\begin{abstract}
Silicon modulators are key components in silicon photonics to support the dense integration of electro-optic (EO) functional elements on a compact chip for various applications including high-speed data transmission, signal processing, and photonic computing. Despite numerous advances in promoting the operation speed of silicon modulators, a bandwidth ceiling of 67 GHz emerges in practices and becomes an obstacle to paving silicon photonics toward Tbps level data throughput on a single chip. Here, we theoretically propose and experimentally demonstrate a design strategy for silicon modulators by employing the slow light effect, which shatters the present bandwidth ceiling of silicon modulators and pushes its limit beyond 110 GHz in a small footprint. The proposed silicon modulator is built on a coupled-resonator optical waveguide (CROW) architecture, in which a set of Bragg gratings are appropriately cascaded to give rise to a slow light effect. By comprehensively balancing a series of merits including the group index, photon lifetime, electrical bandwidth, and losses, we found the modulators can benefit from the slow light for better modulation efficiency and compact size while remaining their bandwidth sufficiently high to support ultra-high-speed data transmission. Consequently, we realize a modulator with an EO bandwidth of 110 GHz in a length of 124 \textmu m, and demonstrate a data rate beyond 110 Gbps by applying simple on-off keying modulation for a DSP-free operation. Our work proves that silicon modulators beyond 110 GHz are feasible, thus
shedding light on the potentials of silicon photonics in ultra-high-bandwidth applications such as data communication, optical interconnection, and photonic machine learning.
\end{abstract}

\section{Introduction}
Silicon modulators are the essential building blocks for silicon photonics\cite{intelnature2004} since they are responsible for converting electrical signals to optical ones which are indispensable in any information technology (IT) applications such as data transmission, interconnection, processing, and computing\cite{datareview2018,zhang2019integrated,feng2011nonreciprocal,samkharadze2018strong}. Leveraged by the CMOS-compatible feature of silicon photonics\cite{rogers2021universal,zhang2022large,shu2022microcomb}, silicon modulators are expected to support wafer-scale manufacturing, massive production, and low cost, thus they are particularly promising in realizing next-generation optoelectronic applications in which an aggregate data throughput of Tbps level is required on a single chip\cite{ducournau2018silicon}, and therefore, modulators with ultra-high-bandwidth and compact size are dispensable in achieving a lane speed above 100 Gbps\cite{reednp2010,reednano2014,like2020,ap2021}. However, in spite of extensive efforts, such as material engineering\cite{gesinp2008,gesimod2018}, device optimization\cite{patelmod2015,simardmod2016,xiaoxipr2018}, and applying new photonic structures\cite{xiaoxiptl2013,zljpr2022}, have paid to promote the operation speed of silicon modulators, their electro-optic (EO) bandwidth has been hindered at a maximum value of 67 GHz to date as a bandwidth ceiling\cite{xiaoxipr2020,ofc67G2022,ol67G2022}, which raises concerns about the potential of silicon photonic toward ultra-high-bandwidth applications. Especially, when looking at the advances that equal or higher bandwidth had been demonstrated upon heterogeneous materials\cite{liu2011graphene,sorianello2018graphene,mosnp2017} on silicon such as lithium niobate\cite{wang2018integrated,he2019high,caixlnc2020,niobite2021,niobite2022}, polymer\cite{lee2002broadband,lugwnc2020,polymer2014,polymer2018}, and plasmonics materials\cite{plasmonic2014,plasmonic2015,plasmonic2018,plasmonic2019,ayata2017high}, it raises a question that whether hybrid integration of a different material on silicon is a necessary approach for high-speed modulation, even in the cost of complex and CMOS-incompatible processes being involved. On the other hand, reducing the device dimension is also critical for dense photonic integration\cite{xu2005micrometre,oecav2007}. Although utilizing resonant or slow light effects\cite{vlasov2005active,baba2008slow} can dramatically improve the modulation efficiency to enable compact size, it also brings a longer photon lifetime that may limit the bandwidth\cite{sunjie2019,jafari2020,tbaba2021}. Therefore, how to realize pure silicon modulators with ultra-high bandwidth and compact footprint under CMOS-compatible manufacturing processes remains an important but elusive problem. 

Here, we theoretically propose a strategy for realizing ultra-high-speed silicon modulators operating at telecom wavelength around 1550 nm by utilizing the slow light effect in a coupled-resonator optical waveguide (CROW) architecture that is consisted of a series of cascaded Bragg gratings. By comprehensively balancing several factors including the group index, photon lifetime, electrical bandwidth, and losses, we design and fabricate the silicon modulators using the standard silicon photonic foundry workflow and demonstrate an ultra-high-bandwidth beyond 110 GHz in a length of 124 \textmu m. To the best of our knowledge, this is a record-high bandwidth among the reported results, which breaks the bandwidth ceiling of silicon modulators to date. As confirmed by the transmission experiment, such a modulator is capable of supporting a data rate beyond 110 Gbps in a spectral window of 8 nm by applying simple on-off keying (OOK) modulation without applying any digital signal processing (DSP). Our findings pave the path for data communication beyond 110 Gps per lane by using silicon photonics, and importantly, highlight the potential of silicon photonics toward ultra-high-bandwidth applications.

\section{Design and principles}
\setcounter{equation}{0}
\renewcommand{\theequation}{\arabic{equation}}

The goal of this work is to develop a pure silicon modulator by utilizing the plasma dispersion effect\cite{soref1987} to achieve EO bandwidth in 100 GHz grade while keeping its footprint as small as possible. Therefore, the target modulator can support data transmission with a single lane speed above 100 Gbps by using simply on-off keying (OOK) coding to best reduce the complexity and cost of DSP for short-range optic links, and reserving the bandwidth budget for even higher lane speed by adopting high order modulation format in the future. For this purpose, we found a carefully designed slow light effect induced from a CROW architecture that can effectively enhance the modulation efficiency while remaining the speed of response doesn't degrade.

In general,  the EO bandwidths of silicon modulators ($BW_{EO}$) are determined together by their optical bandwidth ($BW_{O}$) which is characterized by the photon lifetimes (or equivalently, the quality factors $Q$s) of optical modes, and their electrical bandwidth ($BW_{E}$) that is governed by the RC constants of PN junctions and metal contacts, followed by\cite{jafari2019high}:
\begin{equation}
    \frac{1}{BW_{EO}^2} =\frac{1}{BW_{E}^2} +\frac{1}{BW_{O}^2} .
\label{equation:feo}
\end{equation}
In order to promote the overall bandwidth of modulators, it is essential to improve the optical and electrical bandwidths at the same time and find an appropriate balance between them.

We start with a schematic design of a fish-bone-like slow light waveguide as shown in Fig.~1a, which consists of a series of cascaded Bragg grating and acts as one arm of the silicon Mach-Zehnder modulator (MZM). Specifically, the light propagates along the waveguide in $x$ direction, and the unit cell of the grating is composed of two fingers protruding in $y$ direction with a lattice constant of $a$ along $x$ direction. Further, each number-of-period ($N_{p}$) of unit cells forms one arm of a resonator and the two arms are separated by a $\lambda/4$ phase shifter region from their adjacent one. Therefore, the slow waveguide can be understood as a typical CROW with a finite number of cascaded resonators, counted by number-of-resonator ($N_{r}$).

First, we investigate the optical characteristics of the slow light waveguide. For the mentioned infinite grating, its bulk bands in the vicinity of $X$-point, namely near $ka/2\pi=0.5$, are presented in Fig~1b, in which the anti-symmetric and symmetric transverse electric (TE) bands are denoted as TE-A and TE-B with a bandgap $\Delta$ between them. The bandgap $\Delta$ depicts the coupling strengths between the counter-propagating waves in the grating and thus can be tuned by the geometry of the unit cell. When a finite number of unit cells are cascaded through the phase shifter region, it creates a CROW supercell in the size of 2$N_p$.  Accordingly, TE-A and B bands split into a set of discrete modes as reported\cite{liang2011three,liang2012three,yang2014analytical, wang2016mode,jin2019topologically,chen2022analytical,chen2022observation}, and we plot the lowest-order modes in Fig.~1c for TE-B (upper panel) and TE-A (lower panel) bands for examples. More importantly, the phase shifter gives rise to a mid-gap mode embedded in the bandgap (red dot, Fig.~1b) with its electromagnetic field mostly localizing in the phase shifter region (mid panel, Fig.~1c). As discussed in the literature\cite{gao2020dirac}, such mid-gap modes in phase-shifted gratings are in fact topological and are mathematically equivalent to the Jackiw–Rebbi zero mode\cite{jackiw1976solitons}.

From the view of CROW\cite{yariv1999coupled,poon2004designing}, both the discretized modes of bulk bands and mid-gap mode evanescently couple to the neighbor supercell in coupling coefficients $\kappa =\int d^3 r[\varepsilon_0 (r)-\varepsilon (r)]E_\Omega (r)\cdot E_\Omega (r-2N_p a)$, in which $E_\Omega$ is the normalized electric field of modes. As a result, the couplings bring several continuous CROW bands as we presented in Fig.~1d which clearly exhibit the slow light effect, measured by the group velocity $v_g=d\omega/dk$. Further, the CROW bands open several transmission windows in the spectrum as illustrated in Fig~1e, and the spectrum width follows $\Delta\lambda \propto 2\kappa \lambda_0$, in which $\lambda_0$ is the central wavelength of the modes. We calculate the group index $n_g=v_g/c$ for the CROW band of mid-gap mode (Fig.~1f) and find the group indices are keeping a relatively constant value of $n_g=6.1$ near the center of the transmission window, which is suitable for the silicon modulator to operate thus we call it ``passband". Noteworthy to mention that, to avoid the transmission window being too narrow, the group index $n_g$ cannot be too high, namely the coupling coefficient $\kappa$ cannot be too small.

To elaborate on the relationship and constraint between the design parameters, we simplify the coupling coefficient $\kappa$ from evaluating the tunneling of the mid-gap mode through the bandgap, as:
\begin{equation}
    \kappa =\kappa_0 e^{-2N_p a \cdot \Delta } /N_p 
\label{equation:kappa2}
\end{equation}
where $\kappa_0$ is a constant determined by the unit cell structure; the gap depth $\Delta$ presented how strong the field is bound in the phase shifter region and $2N_p a$ depicts the distance to tunneling through (see Supplementary Section 1 for the details). Then the group index at the transmission window center can be explicitly written as  $n_{g} ={\lambda_0}/{4\pi \kappa N_p a }={(\lambda_0/{4\pi \kappa_0 a } )e^{2N_p a \cdot \Delta}}$ which is exponentially increasing with $\Delta$ and $N_p$, indicating a stronger slow light effect.

It is well known that slow light is beneficial in enhancing the modulation efficiency of the silicon modulators that utilize the plasma dispersion effect\cite{baba2008slow,npslow2008,passoni2020optimizing,hosseini2018energy}. The modulation in carrier concentration causes electrically controlled refractive index changes in silicon materials, resulting in an observable phase accumulation when light travels through. The details of modulators including the geometry parameters, doping level, and contact design are presented in Fig~2a. In such a device, the phase accumulated in a length of $L=2N_r N_p a$ is given by:
\begin{equation}
    \phi = \Delta n_\text{eff} \cdot L_\text{eff} \cdot \frac{2\pi}{\lambda_0} 
    =\eta \cdot n_g \cdot  (2N_r N_p a) \cdot \frac{2\pi}{\lambda_0 } 
\label{equation:phi}
\end{equation}
where $\eta=\eta_0(\sqrt{V+V_{bi}}-\sqrt{V_{bi}}) $ with $\eta_0$ being a coefficient related to the overlapping between the carriers and optic field; $V_{bi}$ is the build-in voltage of the PN junction and $V$ is the operating voltage of the modulator. Accordingly, we define the efficiency factor as $\phi/\pi V$ to characterize the phase change per voltage. The phase accumulation is enhanced by the slow light effect in a factor of $n_g$. Given a determined $n_g$ from $\Delta$ and $N_p$, the modulation efficiency linearly increases with $N_r$, but shows square-root dependency on the voltage (see Supplementary Section 2 for the details).

We argue that the optical bandwidth ($BW_{O}$) sets an upper bound of $n_g$ and $N_r$. Specifically, 
the optical bandwidth can be depicted by the $Q$ of mid-gap mode, given by $BW_{O}=c/(\lambda_0 Q)$, in which the $Q$ are contributed by two parts as $Q^{-1}=Q^{-1}_\text{loading}+Q^{-1}_\text{propagating}$. Here $Q_\text{loading}$ stands for the energy leakage at the ends of the slow waveguide, while  $Q_\text{propagating}$ counts for the energy dissipation when the light propagates through the waveguide due to scattering loss and material absorption. Notice that $Q_\text{propagating}$ remains as a constant value with increasing waveguide length, and we estimate $Q_\text{propagating } \sim 5000$ from simulation. Moreover, $Q_\text{loading}$ can be understood from the proportion between the total energy $W$ stored in the waveguide and the energy flux $P_\text{loading}$ escaping from its ends. We calculate the maps of efficiency factor, $Q$s, and optical bandwidth on varying $N_p$ and $N_r$ (see Supplementary Section 3 for the details ), as illustrated in Fig.~1b to d. Noteworthy that there is a trade-off between modulation efficiency and optical bandwidth. To guarantee the high optical bandwidth to support a silicon modulator beyond 110 GHz, we found $BW_{O}\sim $ 140 GHz is an appropriate value, corresponding to $n_g=6.1$ and efficiency factor of $0.013~\pi/V$ (star marks).

We further discuss the electrical bandwidth ($BW_{E}$) and modulation efficiency. Firstly the doping level of the PN junction would influence such two attributes in a contrary way. As presented in Fig.~2e and f, a higher doping concentration would result in a larger efficiency factor under the same voltage, but because the parasitic resistance and capacitance increase with heavier doping, the electrical bandwidth decrease accordingly. Therefore, we choose a relatively lower doping level of $5.0 \times 10^{17}$/cm$^3$ for our design, which gives $BW_{E} \sim$ 200 GHz at 4 V voltage, to support the overall $BW_{EO}$ beyond 110 GHz.

An operational modulator must accumulate a large enough phase to achieve sufficient intensity modulation depth through Mach-Zehnder interference. As shown in Eq.~3, it seems can be realized by simply lengthening the CROW with large $N_r$. However, other factors including the phase-mismatching between optical and microwave velocities, optical loss of the CROW, and electrical loss of the metal contacts would set an upper limit of the total length. As discussed in Supplementary Section 4, we found that the loss of contacts is the bottleneck among many factors. We calculated the normalized $S_{21}$ responses for a series of contact lengths by applying the data estimated from our previous experiments (Fig.~2g). We found when the contact length is longer than 200 $\mu$m, the electrical loss dramatically increases to an unacceptable level when operating at 200 GHz. Therefore, the limit length of the modulator results in a lower bound of $n_g$, namely the slow light effect cannot be too weak. Through validating the mentioned constraints, we confirmed a modest $n_g =6.1$ is sufficient to accumulate an operational phase in a length $\sim$ 150 $\mu$m.

To summarize the design strategy, the key to ultra-high-speed silicon modulators is to find an appropriate group index $n_g$ by designing the unit cell geometry of CROW, which opens a wide enough transmission window of mid-gap mode but doesn't limit the optical bandwidth to coordinate with electrical bandwidth; On the other hand, sufficient large  $n_g$ is required to accumulate phases in a limit length. Therefore, a silicon modulator beyond 110 GHz is feasible by comprehensively balancing the mentioned factors.

\section{Sample fabrication and experimental characterization}

The proposed strategy offers a guideline to find the optimal design parameters of a silicon EO modulator working under the CROW scheme, with a compromise in high bandwidth and efficiency. To verify the theory, a silicon modulator targeting a bandwidth of $>$100 GHz is fabricated from commercial silicon photonic foundry. The grating period is designed to be 300 nm, with a side wall corrugation depth of 190 nm. The $N_p$ and $N_r$ are selected as 20 and 10 respectively (starred point in Fig. 2b-d). All the above feature sizes are feasible for the standard foundry-based process (More fabrication details could be found in Methods). The visual morphology of the fabricated device under an optical micrograph is shown in Fig. 3a. Compared with the conventional silicon modulator, the footprint is much more compact which is over an order of magnitude shorter (in a length of $\sim$ 124 \textmu m). Fig.~3b illustrates the scanning electron microscope (SEM) picture of the zoom-in top view of the fabricated CROW, showing that the grating teeth become slightly rounded due to fabrication imperfection which is tolerable for the designed performance.

To characterize the performance of the fabricated silicon modulator, we construct a measurement system as schematically shown in Fig.~3c. A tunable continuous wave (CW) laser incidence and radio-frequency (RF) signal are applied to the silicon modulator to drive the modulator for measuring the optical and electrical bandwidth. For evaluating the performance of high-speed data transmission, an oscilloscope is employed to monitor and collect the data, to quantitatively obtain the optical modulation amplitude (OMA), signal-to-noise ratio (SNR), and extinction ratio (ER) of the optical link. See Supplementary Section 5-7 for the details of specific measurements.

Fig. 4a shows the measured optical spectrum of the modulator around 1.55 \textmu m, in which several transmission windows of mid-gap mode and other discretized modes of bulk bands are observed. Large attenuation on both sides of the spectrum is mainly due to the bandwidth limitation of the grating couplers we adopted as input/output optical ports. The passband (shading area of mid-gap mode) around 1550 nm holds an out-of-band rejection ratio $>$60 dB in a flat optical window width of $\sim$8 nm  (Fig. 4b), showing great consistency with the design (red lines). Importantly, such a device exhibits a much larger EO bandwidth than ever before among those reported pure silicon modulators. As shown in Fig.~4c, the frequency response is flat over the entire 110 GHz measurement range, without fast roll-off of bandwidth degrading. The 3 dB EO bandwidth is read to be about 110 GHz from fitting the frequency response curve, which is slightly degraded due to the intrinsic loss of the RF probe. Moreover, the remarkably high electrical bandwidth helps to eliminate the signal impairment when applying an ultra-high-speed driving signal, especially for the serial modulation format such as OOK, in which the Nyquist bandwidth is doubled compared with its PAM4 counterpart. Therefore, a 112 Gbps OOK eye-diagram could be real-time recovered without any pre-equalization at the transmitter side (See Fig.~4d), bringing the benefits in a reduction of the latency budget, as well as lowering the power consumption of the digital signal processing.

The compact footprint, the large EO bandwidth, and together with the wide optical operation window, make the modulator suitable for DWDM-based optical communication applications. As proof, we perform the high-speed optical signal generation at a series of wavelengths within the 8 nm passband, as shown in Fig.~4f. The results show great consistency in each wavelength, confirmed by the measured variations in OMA, SNR, and ER which are smaller than 0.5 dB across the passband. 

We then perform the bit error rate (BER) measurement under different data rates. As shown in Fig.~4g, we measured the BER at different working wavelengths, with the OOK signal speed set as 70 Gbps, 84 Gbps, 98 Gbps, and 112 Gbps. A slight performance fluctuation could be found, which may be due to the non-flatten gain spectrum of the pre-amplifier at receiving end.  Fig~4h further illustrates the BER curves under the same data rates with the variations of the receiving power. With a proper SNR, the BERs can drop well below the hard-decision forward-error-coding (HD-FEC) threshold (3.8 $\times$ 10$^{-3}$) under the data rate of 98 Gbps, and below the soft-decision forward-error-coding (SD-FEC) threshold (2 $\times$ 10$^{-3}$) when up to 112 Gbps.

\section{Discussions}
The demonstration of a CROW silicon modulator with an unprecedentedly high EO bandwidth of 110 GHz and an ultra-compact length of 124 \textmu m may broaden the horizon of silicon photonics. Both the bandwidth and the footprint of the device, to the best of our knowledge, break the performance limitation of a pure silicon modulator in traditional cognition, without additional complicated processes or aids from heterogeneous functional materials. Also, the work performed the first DSP-free data transmission experiments with a symbol rate beyond 110 GBaud, showing its potential in the next-generation datacom and telecom applications with the requirements of low latency and power consumption. Compared with the other type of resonant modulator - microring/disk modulator\cite{xu2005micrometre,sunjie2019,microdisk2011}, an optimal CROW scheme exhibits a compact size in the same scale which is more comparable in size to their electronic supporting elements, meanwhile holding wider optical bandwidth and being less sensitive to temperature variation, which helps make full use of the spectral resources and save the power budget, benefiting the applications such as optical neural network and highly parallel data transmission, for high volume throughput and large-scale integration\cite{datareview2018}.

Following the proposed guideline, the silicon modulator could be flexibly designed to target different specifications and scenarios. In our high-speed demonstrations, data transmission in OOK format is performed to verify the DSP-free characteristic of large EO bandwidth. Other advanced modulation formats, such as PAM-4 and 16 QAM, could further be employed with an increased modulation depth (e.g. with a larger resonator number $N_r$) for better separation of the multi-level signals, in which cases the aggregation rate of a single carrier could potentially raise up above 1 Tbps. Besides the CROW structure, the high-frequency electrode is another degree of freedom for high-speed design which has not been specifically optimized in this work. Therefore, if the RF loss is further reduced by the substrate removed process\cite{xiaoxipr2018} or the relative match condition is delicately managed\cite{microelectrode2021}, the frequency response of the proposed structure is expected to be even higher. Moreover, the proposed design strategy can also be applied to other integrated platforms, such as III-V or LNOI\cite{niobite2021}, where the footprint is a long-term roadblock for high-density integration and a sophisticatedly designed slow light effect would help. 

\section{Conclusion}
To summarize, we propose and implement a design strategy for silicon modulators by employing comprehensively designed slow light effect in CROW architecture to achieve an EO bandwidth beyond 110 GHz in a compact length of 124 \textmu m. The group index of cascaded mid-gap mode is optimized to a modest value of $n_g=6.1$ by balancing the passband width, optical bandwidth, modulation efficiency, and best coordinates with electrical bandwidth. By using the fabricated modulator, we experimentally demonstrate a data rate beyond 110 Gbps in an operational window width of 8 nm by applying simple on-off keying modulation, which validates the proposed design strategy. Our work breaks the present bandwidth ceiling of silicon modulators and reveals the great potential of silicon photonics for the next-generation ultra-high speed data transmission, ultra-wide-band signal processing and large-volume photonic computing,

\section{Methods}
\noindent \textbf{Design and fabrication of the devices. } 
The silicon modulator was fabricated on a 200 mm SOI wafer with a silicon-layer thickness of 220 nm and a buried oxide layer thickness of 2 \textmu m using a standard 90 nm lithography SOI process at CompoundTek Pte in a one-to-one 200-mm-wafer run. The width of the waveguide core in the fishbone grating structure is 455 nm, with a 90 nm thick partially etched rib area for carrier doping and metal contact. The concentration of P-type and N-type doping at the waveguide core is $5.0 \times 10^{17}$/cm$^3$ and $5.0 \times 10^{17}$/cm$^3$, respectively. To ensure the waveguide mode adiabatically transit into the CROW mode, waveguide tapers are used to connect the single mode strip waveguide (450 nm) and the phase shifting region. A pair of C-band grating couplers are employed for light input/output. As for the electrical part, the modulator is designed to work under the push-pull driving configuration. Thus, a GSGSG-type high-frequency electrode is employed here. To realize the characteristic impedance match-up, the gap between the ground and signal electrode is 6.4 \textmu m, with the Cu electrode thickness of 1.2 \textmu m.  
 
 \vspace{3pt}
\noindent\textbf{Experimental details.}
The optical loss of the modulator was extracted from the power difference between the coupling fiber, excluding an unoptimized coupling loss of $\sim$ 10 dB. The insertion loss of the 124-\textmu m modulator is then measured to be 6.8 dB, incorporating 5.4 dB loss from the phase shifter and remaining from other structures (e.g. 3 dB couplers, routing waveguides, etc). Although the loss per unit length is a bit larger, it is still comparable to a standard 3-mm depletion mode silicon modulator that fits the requirement for real-world deployment.

As for the bandwidth test, a vector network analyzer (Keysight PNA-X Network Analyzer N5247B) with its plug-in 110 GHz lightwave component analyzer (LCA Optical Receiver N4372-67985) is used to measure the linear electro-optic transmission. The connection RF wires hold an electrical bandwidth of 100GHz and are pre-calibrated over 110 GHz before the test. In the data transmission experiments, the high-speed OOK signal with a standard PRBS pattern is generated by a bit pattern generator (SHF 12104A) and then the signal speed is doubled to over 100 Gbaud by a MUX (SHF C603B). At the receiving end, eye diagrams are produced by a sampling oscilloscope (Tektronix DSA8300). For the bit error ratio measurement, the same configuration but an arbitrary wave generator (Keysight M8199A) is adopted for programmable signal generation. After data receiving, the signals are fast sampled by a real-time oscilloscope (Keysight UXR0594AP) and afterward offline processed with Feed-Forward Equalization (FFE) and Maximum-Likelihood Sequence Estimation (MLSE) algorithms. The receiving power is adjusted by a variable optical attenuator before the pre-amplifier (AEDFA-PA-35-B-FA).

\noindent\textbf{Conflict of interest}

\noindent The authors declare that they have no conflict of interest.

\noindent\textbf{Acknowledgments}

\noindent This work was partly supported by National Key Research and Development Program of China (2022YFB2803700, 2021YFB2800400, 2022YFA1404804); National Natural Science Foundation of China under Grant (62235002, 62001010, 12204021, 62135001); Beijing Municipal Science \& Technology Commission
(Z221100006722003); Beijing Municipal Natural Science Foundation (Z210004), and  Major Key Project of PCL (PCL2021A14).
 
\noindent\textbf{Author contributions}
\noindent The experiments were conceived by C.H. The devices were designed by C.H. The slow light theory was developed by Z.Z. The simulations were conducted by C.H. and Z.Z. The system experiments were performed by C.H., with assistance from M.J., H.S., J.Q., Y.T., B.B., and F.Y. The device characterization was conducted by C.H., F.W., and Z.X.Z. The results were analyzed by C.H., Z.Z., H.S., R.C., and B.S. The figure optimization was conducted by C.H., B.S., Y.W., and H.W. All authors participated in the writing of the manuscript. The project was supervised by H.S., S.Y., C.P., and X.W.
\clearpage
\noindent\textbf{References}

\bibliography{Reference.bib}{}
\bibliographystyle{naturemag}

\clearpage
\begin{figure}
\centering
\includegraphics[width=18cm]{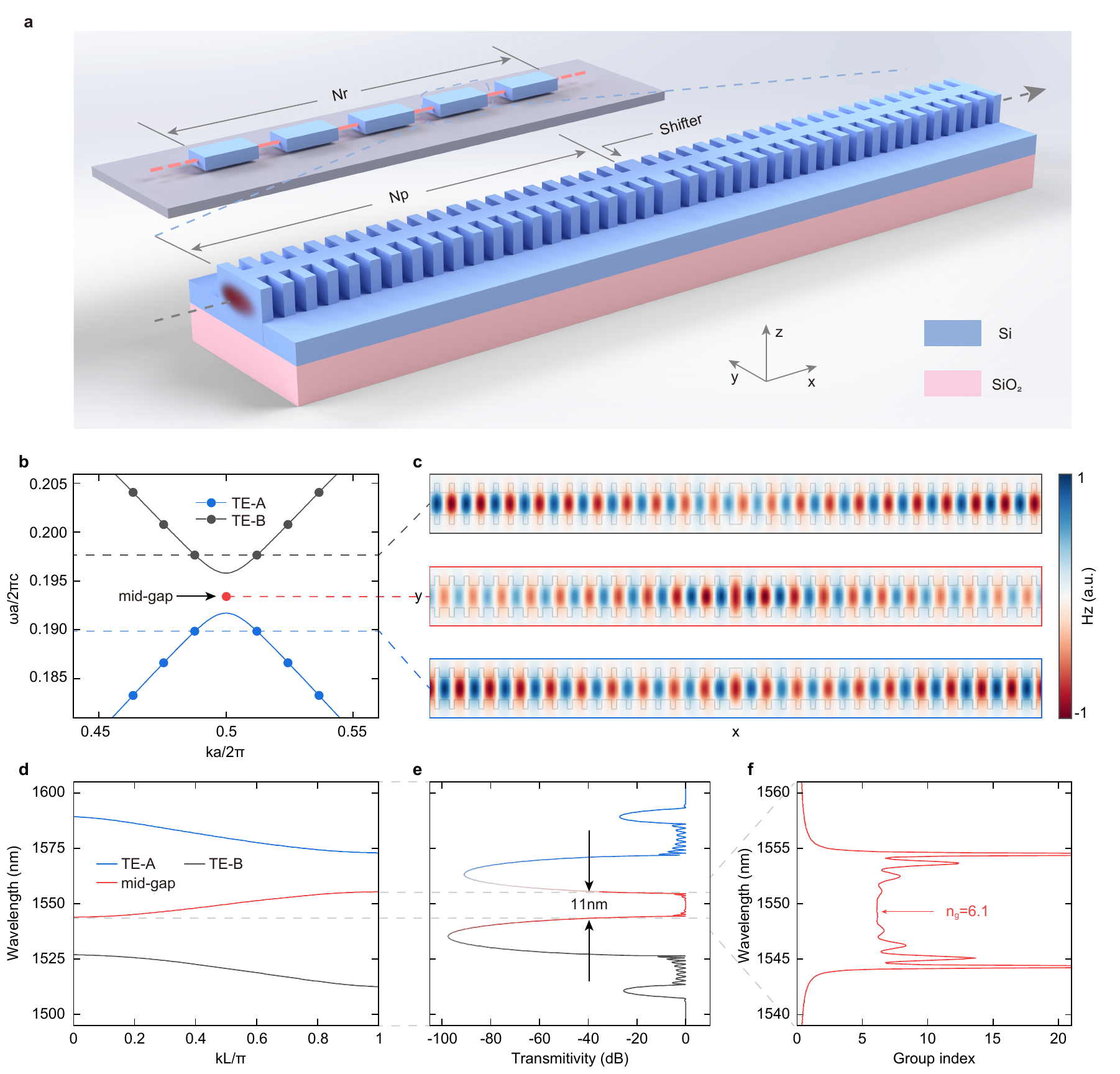}
\end{figure}
\begin{figure}
\centering
\caption{\textbf{Slow light effect in cascade Bragg gratings}.
\label{mfig1}
(a) 
A schematic of the slow light modulator on SOI wafer that consists of a series of fishbone-like Bragg gratings separated by phase shifter regions to form a CROW architecture.
(b) The bulky band structure of the grating shows two bands TE-A and B in the vicinity of $X$ point. When two neighboring gratings couple through the phase shifter, a topologically stable mid-gap mode emerges in the band gap. 
(c)
The magnetic field distributions ($H_z$) of lowest order discretized bulk modes of TE-A (lower) and TE-B (upper), and the mid-gap mode (middle) in the CROW supercell.
(d)
The dispersions of three CROW bands correspond to modes TE-A, TE-B, and the mid-gap mode.
(e)
The transmission spectra of the CROW bands TE-A, TE-B, and the mid-gap mode, showing each of them open a transmission window. The spectrum width associated with the mid-gap CROW band is 11 nm.
(f)
The group index of the mid-gap CROW band indicates a modest slow light effect of  $n_g=6.1$ at the center of the transmission window.
}
\end{figure}

\clearpage
\begin{figure}
\centering
\includegraphics[width=18cm]{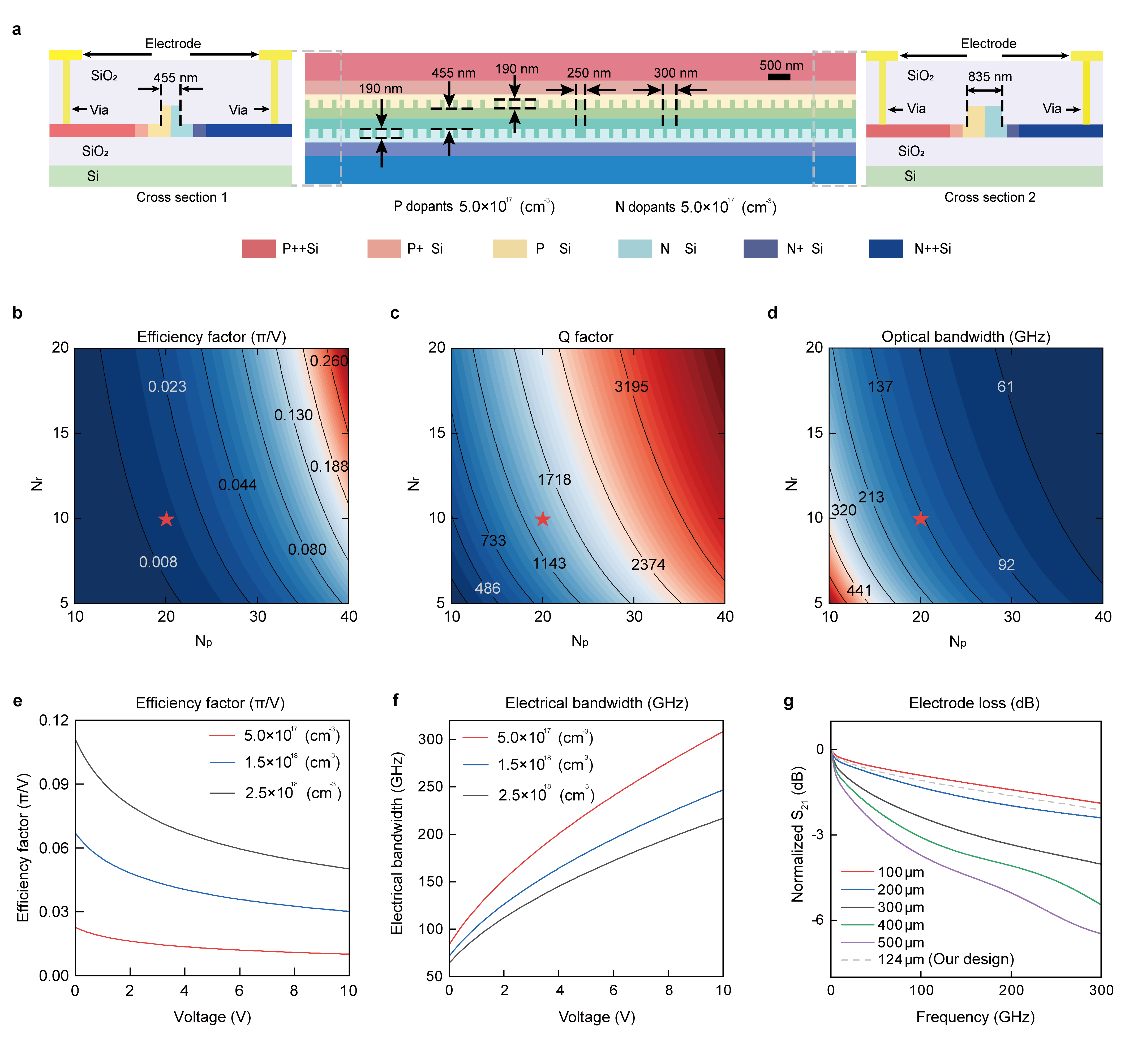}
\caption{\textbf{Design of slow light silicon modulator with ultra-high bandwidth and compact size}.
\label{mfig1}
(a) 
The detailed device geometry and doping configuration of the slow silicon modulator.
(b, c, d) The maps of merits related to optical characteristics upon $N_p$ and $N_r$, for (b) efficiency factor, (c) Q factors, and (d) optical bandwidth.  (e, f) The electrical merits of the silicon modulator under different doping levels, for (e) efficiency factor and (f) electrical bandwidth. 
(g)
Normalized electrode loss at different modulation frequencies and device lengths, indicating a dramatic loss increment when the device length exceeds 200 \textmu m. The star marks show our optimal design.
}
\end{figure}

\clearpage
\begin{figure}
\centering
\includegraphics[width=16cm]{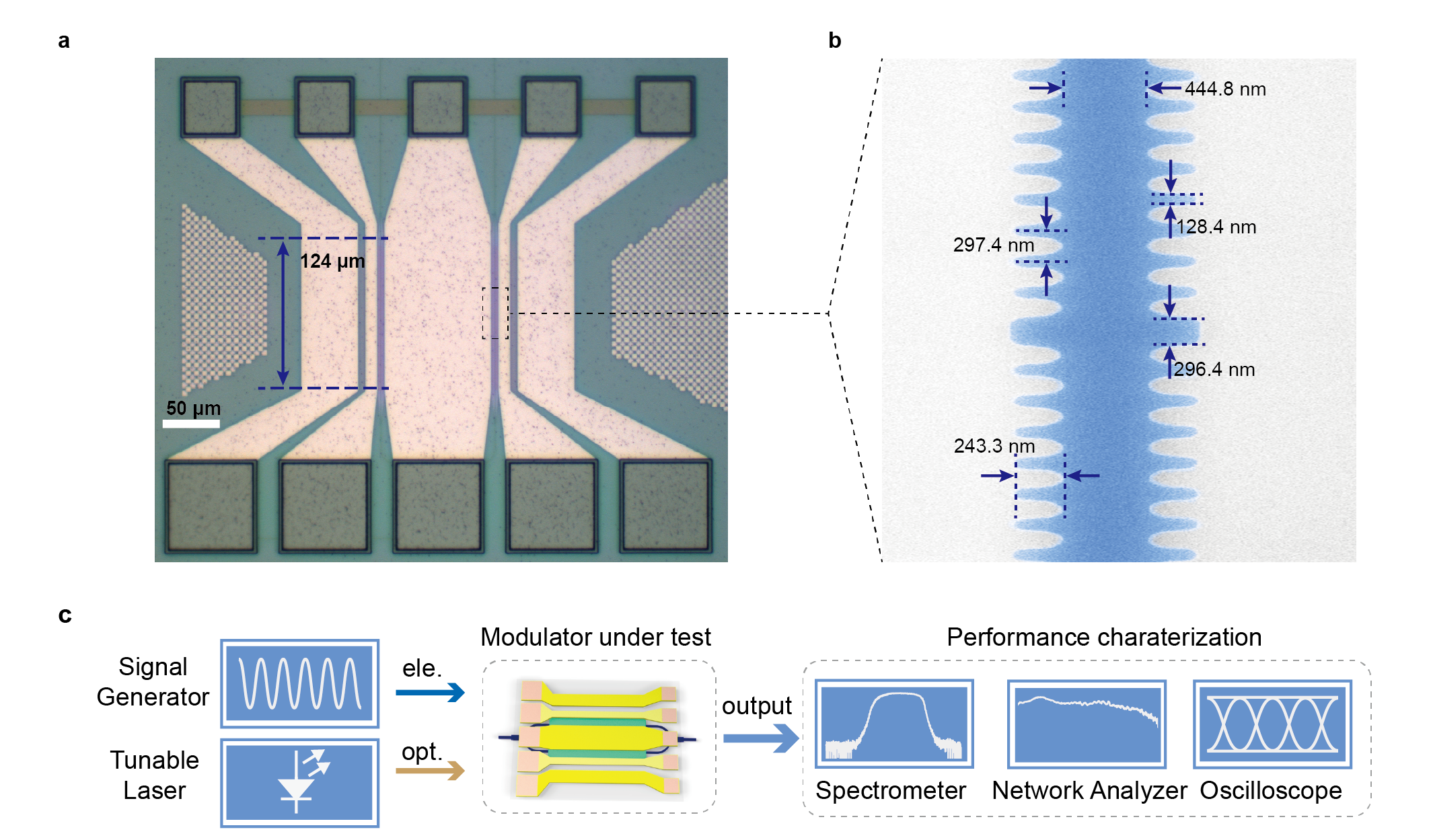}
\caption{\textbf{Fabrication and measurement system of the designed slow light silicon modulator}.
\label{mfig1}
(a) 
Optical micrograph of the fabricated slow light silicon modulator. The modulation region has an ultra-compact footprint, consisting of two 124-\textmu m long, PN-doped CROW arms to form the MZI configuration. 
(b)
The zoom-in SEM image shows part of the Bragg grating, which is composed of a $\lambda/4$ phase shifter and tens periods of fishbone-like gratings on both sides. Some critical dimensions of the fabricated device (e.g. the core width of the waveguide, the grating spacing, the width of the phase shifter region, etc.) are measured and marked.
(c)
The schematic of the measurement system to characterize the fabricated slow light silicon modulator.
}
\end{figure}

\clearpage
\begin{figure}
\centering
\includegraphics[width=18cm]{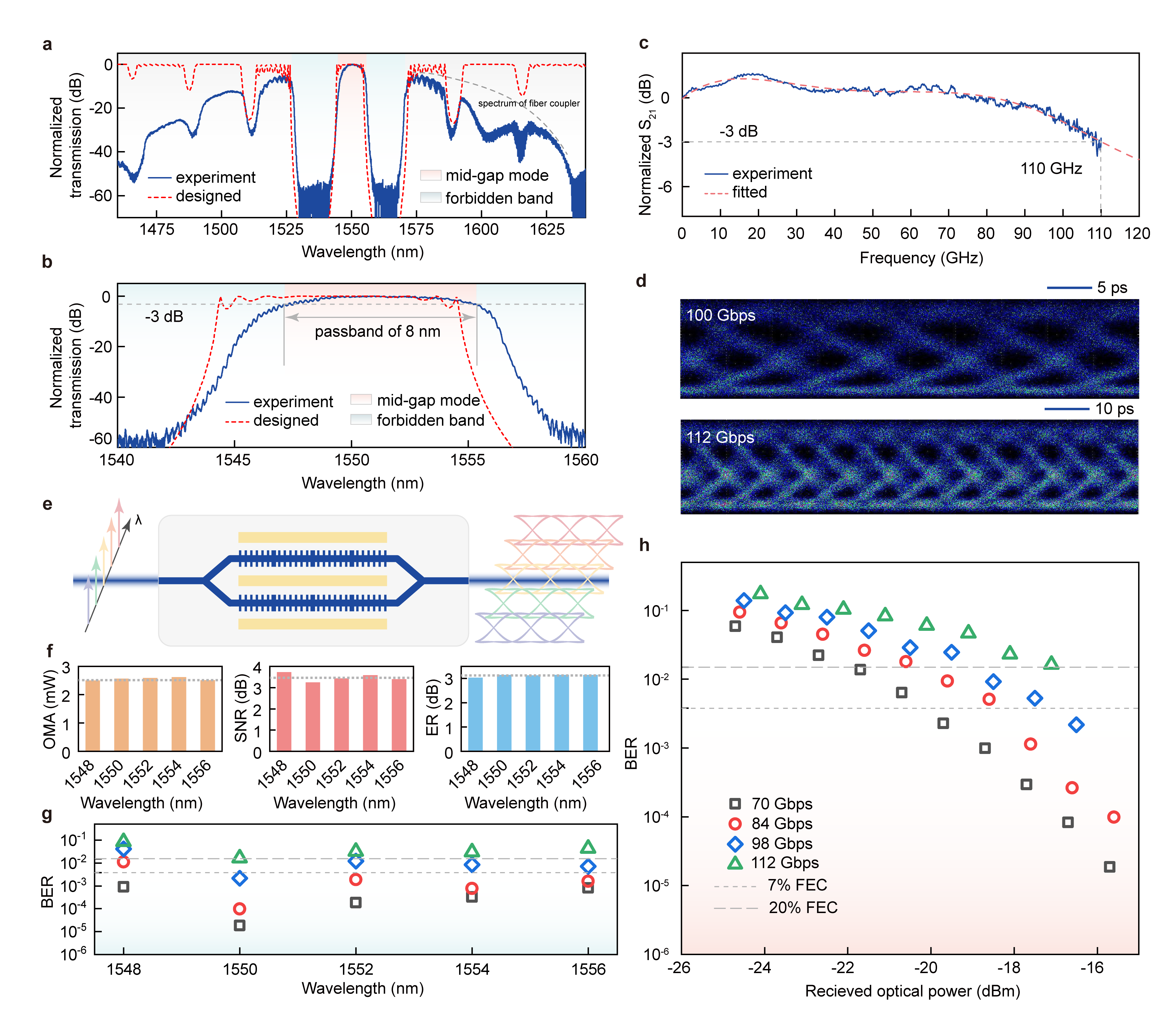}
\end{figure}
\begin{figure}
\centering
\caption{\textbf{Characterization of the ultra-high bandwidth and demonstration ultra-high-speed data transmission}.
\label{mfig1}
(a) The measured transmission spectrum (blue curves) in a range from 1460 nm to 1640 nm, which shows great consistency with the design (red lines).  
(b)
The transmission spectrum of the mid-gap CROW band, namely the passband, resides in the vicinity of 1550 nm. The passband shows as a flat-top window in a spectrum width of 8 nm with $>$ 60 dB out-of-band rejection ratio.
(c)
The measured (blue) and fitted (red) S21 curves of the EO response of the slow light silicon modulator confirm its 3-dB bandwidth reaches 110 GHz.
(d)
The DSP-free optical eye diagrams for the OOK modulation at the data rates of 100 Gbps (upper panel) and 112 Gbps (lower panel), respectively. The measured extinction ratio is 3.15 dB and 2.15 dB, respectively.
(e)
The scheme of the multi-wavelength data transmission using the fabricated slow light silicon modulator. 
(f)
The OMA, SNR, and ER performance across the operational passband of 8 nm.
(g)
The consistency of the measured BERs across multiple wavelengths in the passband under OOK modulation, in data rates of 70 Gbps, 84 Gbps, 98 Gbps, and 112 Gbps.
(h)
The measured BER  versus received optical power under a variety of data rates from 70 to 112 Gbps.
}
\end{figure}
\end{document}